\documentclass[12pt]{article}
\title{\begin{flushright}
{\normalsize McGill/98-43\\NUC-MINN-98/14-T\\}
\end{flushright}
{\bf $J/\psi$ Production and Absorption in High Energy
Proton-Nucleus Collisions}}
\author{{\bf Charles Gale}\thanks{gale@physics.mcgill.ca}\vspace*{0.1in} \\
 {\it Physics Department}\\
 {\it McGill University}\\ \vspace*{0.2in}
 {\it Montreal, Quebec H3A 2T8, Canada}\\ \vspace*{0.1in}
{\bf Sangyong Jeon}\thanks{jeon@nta2.lbl.gov} \\
 {\it Nuclear Science Division}\\
 {\it Lawrence Berkeley National Laboratory} \\ \vspace*{0.2in}
 {\it Berkeley, CA 94720}\\ \vspace*{0.1in}
{\bf Joseph Kapusta}\thanks{kapusta@physics.spa.umn.edu}\\
  {\it School of Physics and Astronomy}\\
  {\it University of Minnesota}\\ {\it Minneapolis, MN 55455}}

\date{}

\usepackage[dvips]{graphicx}
\begin{document}

\maketitle
\begin{abstract}

Measured $J/\psi$ production cross sections for 200 and 450 GeV/c protons
incident on a variety of nuclear targets are analyzed within a Glauber
framework which takes into account energy loss of the beam proton, the time
delay of particle production due to quantum coherence, and absorption of the
$J/\psi$ on nucleons.  The best representation is obtained for a coherence
time of 0.5 fm/c, previously determined by Drell-Yan production in
proton-nucleus collisions,
and an absorption cross section of 3.6 mb, which is
consistent with the value deduced from photoproduction of
the $J/\psi$ on nuclear targets.

\end{abstract}
\noindent
PACS numbers: 13.85.Ni, 24.85.+p, 11.80.La\\

\vspace*{0.1in}


\newpage

Propagation of a highly relativistic particle through a medium is of interest
in several areas of physics.  High energy proton-nucleus scattering has
been studied for many decades by both the nuclear and particle physics
communities \cite{Wit}.  Such studies are particularly relevant for
the Relativistic Heavy Ion Collider (RHIC), which will collide beams
of gold nuclei at an energy of 100 GeV per nucleon, and for the Large
Hadron Collider (LHC), which will collide beams of lead nuclei at
1500 GeV per nucleon \cite{qm97}.

In a recent paper \cite{us} we modeled the production of high mass, Drell-Yan,
pairs of leptons as measured in collisions of 800 GeV/c protons incident on a
variety of nuclear targets by the Fermilab collaboration E772 \cite{E772}.  Our
modeling was done on the basis of hadronic degrees of freedom.  We took into
account the energy loss of the beam proton as it traversed the nucleus as well
as the Landau-Pomeranchuk-Migdal effect \cite{lpm}.  The latter basically
acknowledges that there is a time ordering in the appearance of produced
particles; the hard particles (Drell-Yan pairs) appear before the soft
particles (typically pions).  This is a quantum mechanical effect, essentially
the uncertainty principle.  By fitting to the atomic mass number dependence
of the Drell-Yan cross section at high Feynman $x$ we were able to infer a
value for the proper coherence or formation time of 0.4$\pm$0.1 fm/c. This
value is about what should be expected {\it a priori}.  In the center of mass
frame of the colliding nucleons at the energies of interest a typical pion is
produced with an energy of E$_{\pi} \approx 500$ MeV.  By the uncertainty
principle this takes a time of order $\hbar c/E_{\pi} \approx 0.4$ fm/c.

A related process is the production of $J/\psi$ in high energy proton-nucleus
collisions which we shall address in this paper.  This is also a relatively
hard process and so both energy loss of the beam proton and the
Landau-Pomeranchuk-Migdal effect must be taken into account. However, there
is an additional effect
which plays a role, and that is the occasional absorption or breakup of the
$J/\psi$ in encounters with target nucleons.  (The inelastic interaction of one
of the leptons in Drell-Yan production with target nucleons is ignorably
small.)  The absorption cross section, $\sigma_{\rm abs}$, has been estimated
in a straightforward Glauber analysis without energy loss and with an infinite
coherence/formation time to be about 6-7 mb \cite{DimaJ}.  This has formed the
basis for many analyses of $J/\psi$ suppression in heavy ion collisions.  Any
anomalous suppression may be an indication of the formation of quark-gluon
plasma \cite{qm97,matsui}, hence the importance of obtaining the most accurate
value of $\sigma_{\rm abs}$ possible.  This cross section has also been
inferred from photoproduction experiments of $J/\psi$ on nuclei from which a
value much less than that has been obtained \cite{photo}.  This has been a
puzzle. One attempt to resolve this apparent discrepancy consists of
modeling the produced $J/\Psi$ state as a pre-resonant color dipole state
with two octet charges \cite{dima2}; however, the results are only
semi-quantitative.

For a basic description of high energy proton-nucleus scattering
we prefer to work with hadronic variables rather than partonic ones.
We make a straightforward linear extrapolation from proton-proton
scattering.  This extrapolation, referred to as LEXUS, was detailed
and applied to nucleus-nucleus collisions at beam energies of several
hundred GeV per nucleon in ref. \cite{lexus}, and to Drell-Yan production
in 800
GeV proton-nucleus collisions in ref. \cite{us}.  In order to compute the
production cross section of $J/\psi$ in proton-nucleus collisions we need a
parametrization of it in the more elementary nucleon-nucleon collisions. For
this we call upon the parametrization of a compilation of data by
Louren\c{c}o \cite{Carlos}.
\begin{equation}
B \sigma_{NN \rightarrow J/\psi}(x_F > 0) = 37 \left( 1 - m_{J/\psi}/\sqrt{s}
\right)^{12} \, {\rm nb}
\end{equation}
Here $B$ is the branching ratio into dimuons and $x_F$ is the ratio of the
momentum carried by $J/\psi$ to the beam momentum in the center of mass frame
$(-1 < x_F < 1)$.  Due to the degradation in momentum of the proton as it
traverses the nucleus it is important to know the $x_F$ dependence of the
production.  The Fermilab experiment E789 has measured this dependence at 800
GeV/c \cite{E789} to be proportional to $(1-|x_F|)^5$.  Assuming that this
holds at lower energy too we use the joint $\sqrt{s}$ and $x_F$
functional dependence and magnitude:
\begin{equation}
\frac{d\sigma_{NN \rightarrow J/\psi}}{dx_F} = 6 \sigma_{NN \rightarrow
J/\psi}(x_F > 0) (1-|x_F|)^5 \, .
\end{equation}
The cross section in proton-nucleus collisions can now be computed in LEXUS
with no ambiguity.
\begin{figure}
\begin{center}
\includegraphics[angle=90, width=10cm]{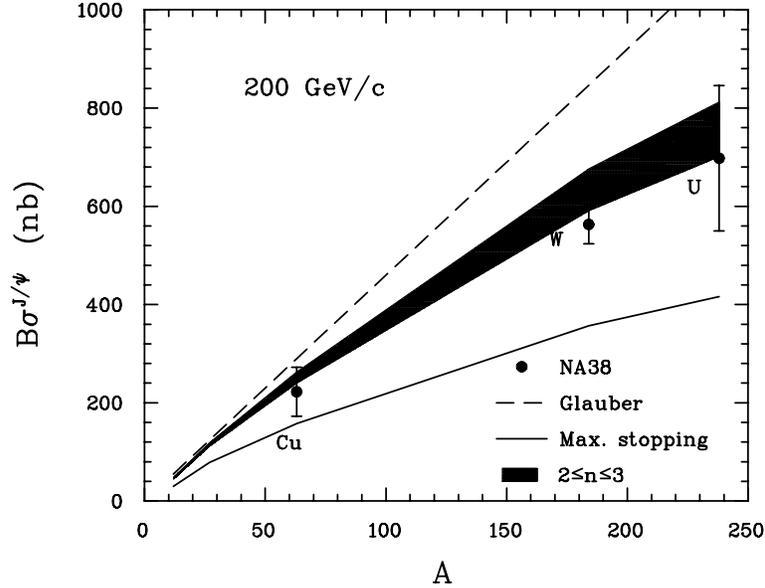}
\end{center}
\caption{\small
Branching ratio into muons times cross section to produce $J/\psi$
with $x_F \ge 0$ in proton-nucleus collisions at 200 GeV/c.  The data is from
NA38 \cite{Carlos}.  The dashed line is $A$ times the nucleon-nucleon
production cross section.  The solid curve represents full energy
loss with zero coherence/formation time, while the banded region represents
partial energy loss
with a coherence/formation time within the limits set by Drell-Yan production.
(Computations were done for C, Al, Cu, W and U and the points connected by
straight lines to guide the eye.)}
\end{figure}

Figures 1 and 2 show the results of our calculation in comparison to data taken
by NA38 \cite{Carlos}.  The dashed curves
are $A$ times the nucleon-nucleon production cross section; they obviously
overestimate the data.  
\begin{figure}
\begin{center}
\includegraphics[angle=90, width=10cm]{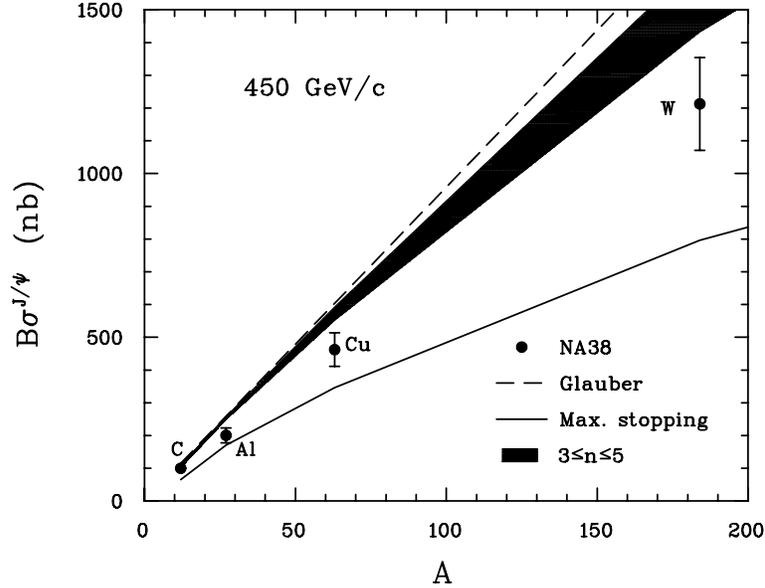}
\end{center}
\caption{\small
 Same as figure 1 but for a beam momentum of 450 GeV/c.}
\end{figure}
The
solid curves show the result of LEXUS with full energy degradation of the beam
proton without account taken of the Landau-Pomeranchuk-Migdal effect; they
obviously underestimate the data.  The hatched regions represent the inclusion
of the latter effect with a proper formation/coherence time $\tau_0$ in the
range of 0.3 to 0.5 fm/c consistent with Drell-Yan production \cite{us}.  The
time delay is implemented as follows \cite{us}:  The energy available for the
production of $J/\psi$ is that which the proton had $n$ collisions prior; that
is, the previous $n$ collisions are ignored for the purpose of determining the
proton's energy. This is an approximate treatment of the
Landau-Pomeranchuk-Migdal effect.  The $n$ is related to the beam energy and
to  the coherence time $\tau_0$ in the center of mass frame of the
colliding nucleons.  The proper coherence time is essentially the same as the
proper formation
time of a pion since most pions are produced with rapidities near
zero in that frame.  The first proton-nucleon collision is the most
important, so boosting this time into the rest frame of the target
nucleus and converting it to a path length (proton moves essentially
at the speed of light) gives $\gamma_{\rm cm}\,c\,\tau_0 \approx
\sqrt{\gamma_{\rm lab}/2}\,c\,\tau_0$.  This path length may then be
equated with $n$ times the mean free path
$l = 1/\sigma^{\rm tot}_{\rm NN} \rho$.  Using a total cross section
of 40 mb, a nuclear matter density of 0.155 nucleons/fm$^3$, and
$0.3 < \tau_0 <
0.5$ fm/c we obtain $2 < n < 3$ at 200 GeV/c and $3 < n < 5$ at 450 GeV/c. As
may be seen from the figures, the data is overestimated, indicating the
necessity for nuclear absorption.

\begin{figure}[!t]
\begin{center}
\includegraphics[angle=90, width=10cm]{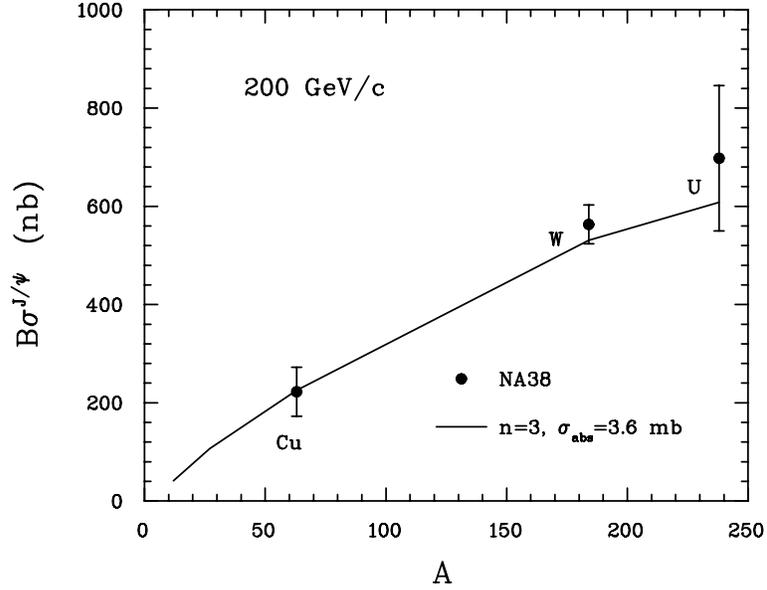}
\end{center}
\caption{\small
Same data as in figure 1.  The solid curve is the best fit of the
model which includes beam energy loss with a coherence time of 0.5 fm/c (n=3 at
this energy) and a $J/\psi$ absorption cross section of 3.6 mb.}
\end{figure}
\begin{figure}[!b]
\begin{center}
\includegraphics[angle=90, width=10cm]{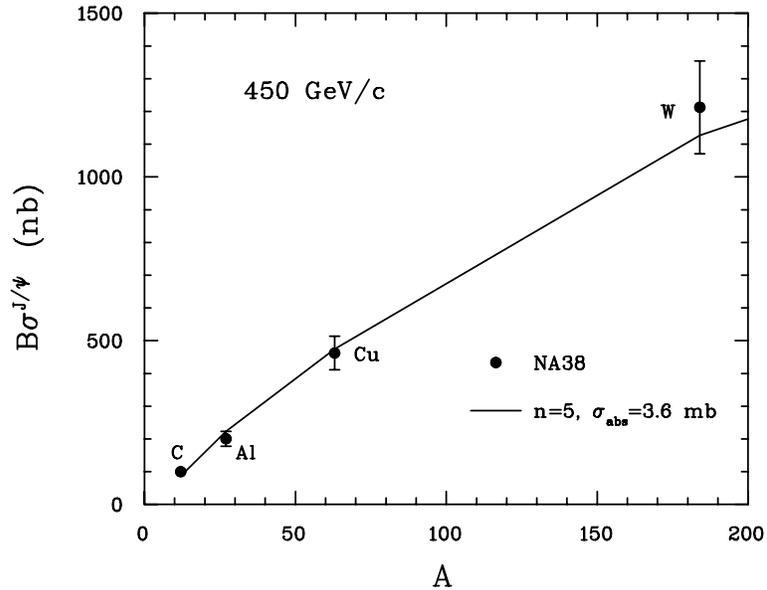}
\end{center}
\caption{\small
Same data as in figure 2.  The solid curve is the best fit of the
model which includes beam energy loss with a coherence time of 0.5 fm/c (n=5 at
this energy) and a $J/\psi$ absorption cross section of 3.6 mb.} 
\end{figure}
We now introduce a $J/\psi$ absorption cross section on nucleons and compute
its effect within LEXUS in the canonical way \cite{DimaJ}.  When the $J/\psi$
is created there will in general be a nonzero number of nucleons blocking its
exit from the nucleus.  Knowing where the $J/\psi$ is created allows one to
calculate how many nucleons lie in its path, and hence, to compute the
probability that it will be dissociated into open charm.  We choose a value
of $\tau_0$ allowed by Drell-Yan measurements, mentioned above, and then
vary $\sigma_{\rm abs}$, assuming that it is energy
independent.  The lowest value of chi-squared for the 200 and 450 GeV/c
data set taken together is obtained with $\tau_0 = 0.5$ fm/c and
$\sigma_{\rm abs} = 3.6$ mb.  The results are shown in figures 3 and 4.
The fitted values all lie within one standard deviation of the data
points.  This is quite a satisfactory representation of the data.  It means
that both Drell-Yan and $J/\psi$ production in high energy proton-nucleus
collisions can be understood in terms of a conventional hadronic
analysis when account is taken of the energy loss of the beam proton, the
Landau-Pomeranchuk-Migdal effect, and nuclear absorption of the $J/\psi$
in the final state.  It also means that the absorption cross section for
$J/\psi$ inferred from high energy proton-nucleus collisions is
consistent with the value inferred from photoproduction experiments on nuclei.

It will be very instructive to repeat this
analysis in the language of partonic variables.  Actually, the analysis
with parton energy loss alone was reported by Gavin and Milana \cite{sean}
with satisfactory results obtained for Drell-Yan and $J/\psi$ if the partons
lose about 1.5 GeV/fm.  Nuclear shadowing \cite{shadows}
needs to be taken into account too.  The
relationship among all these effects is not well-understood, nor is the
relationship between these effects in partonic and hadronic variables.
Finally, the implications for nucleus-nucleus collisions
\cite{qm97} will undoubtedly be important; they are under investigation.

\section*{Acknowledgements}

The authors gratefully acknowledge discussions with S. Gavin, D. Kharzeev, C.
Louren\c{c}o, J. Moss, and H. Satz.
C.G. thanks the School of Physics and Astronomy at the University of
Minnesota for its hospitality during a sabbatical leave.
C.G. and J.K. thank the Institute for Nuclear Theory at the University
of Washington for its hospitality and the Department of Energy for
partial support during the program ``Probes of Dense Matter in
Ultrarelativistic Heavy Ion Collisions".  This work was also supported
by the U. S. Department of Energy under grant
DE-FG02-87ER40328, by the Natural Sciences and Engineering Research Council
of Canada, by the Fonds FCAR of the Quebec Government, and  by the
Director, Office of Energy Research, 
Office of High Energy and Nuclear Physics, Division of Nuclear Physics,
of
the U.S.~Department of Energy under Contract No.~DE-AC03-76SF00098
and DE-FG03-93ER40792.

\end{document}